# RADIO OBSERVATIONS OF THE $\gamma$-RAY BLAZAR 0528+134

M. POHL[1], W. REICH[2], T. P. KRICHBAUM[2], K. STANDKE[2,3],
S. BRITZEN[2], H. P. REUTER[2], P. REICH[2], R. SCHLICKEISER[2],
H. UNGERECHTS[4], R. L. FIEDLER[5], E. B. WALTMAN[5],
K. J. JOHNSTON[6] AND F. D. GHIGO[7]

[1] *MPE, Postfach 1603, 85740 Garching, Germany*
[2] *MPIfR, Postfach 2024, 53010 Bonn, Germany*
[3] *University Bonn, Nussallee 17, 53121 Bonn, Germany*
[4] *IRAM, Ave. Divina Pastora 7, 18012 Granada, Spain*
[5] *NRL, Code 7210, Washington, DC 20375-5351, USA*
[6] *US Naval Obs., Washington, DC 20392-5420, USA*
[7] *NRAO, P.O. Box 2, Green Bank, WV 24944, USA*

## 1. Radio monitoring of 0528+134

We report multifrequency observations of the $\gamma$-ray blazar 0528+134 with the Effelsberg 100-m telescope, the IRAM 30-m telescope at Pico Veleta and the NRL Green Bank Interferometer. The observing methods are described elsewhere (Reich *et al.*, 1993; Pohl *et al.*, 1995). The radio lightcurves are given in Fig.1 in comparison to the status of 0528+134 in the EGRET energy range. The uncertainties in the flux densities quoted there are less than 5% at 10.55 GHz and lower frequencies, while slightly exceeding this value at 32 GHz and 86 GHz.

Fluctuations at 2.25 GHz and 2.695 GHz of about 20% – much larger than the errors – are rather frequent during the time of monitoring and these fluctuations are significantly larger than those at 4.75 GHz, 8.3 GHz and 10.55 GHz. These fluctuations may be explained as stochastic interstellar scattering and are probably not intrinsic. The unusual depression in July 1993 is most likely an extreme scattering event (Pohl *et al.*, 1995).

0528+134 underwent a major radio and mm outburst in 1993 a few month after a very strong outburst in high energy $\gamma$-rays. The evolution of the source during the outburst follows the canonical behaviour of appearing first at high frequencies, and then with decreasing optical depth and energy loss of the radiating particles being detected at lower and lower frequencies. A similar behaviour is indicated for the weaker $\gamma$-ray outburst in May 1991 and the mm activity in the beginning of 1992. This result fits nicely to the general finding that blazars are bright in $\gamma$-rays preferrentially at the beginning of a radio outburst (Valtaoja and Teräsranta, 1995; Mücke et al., 1996).

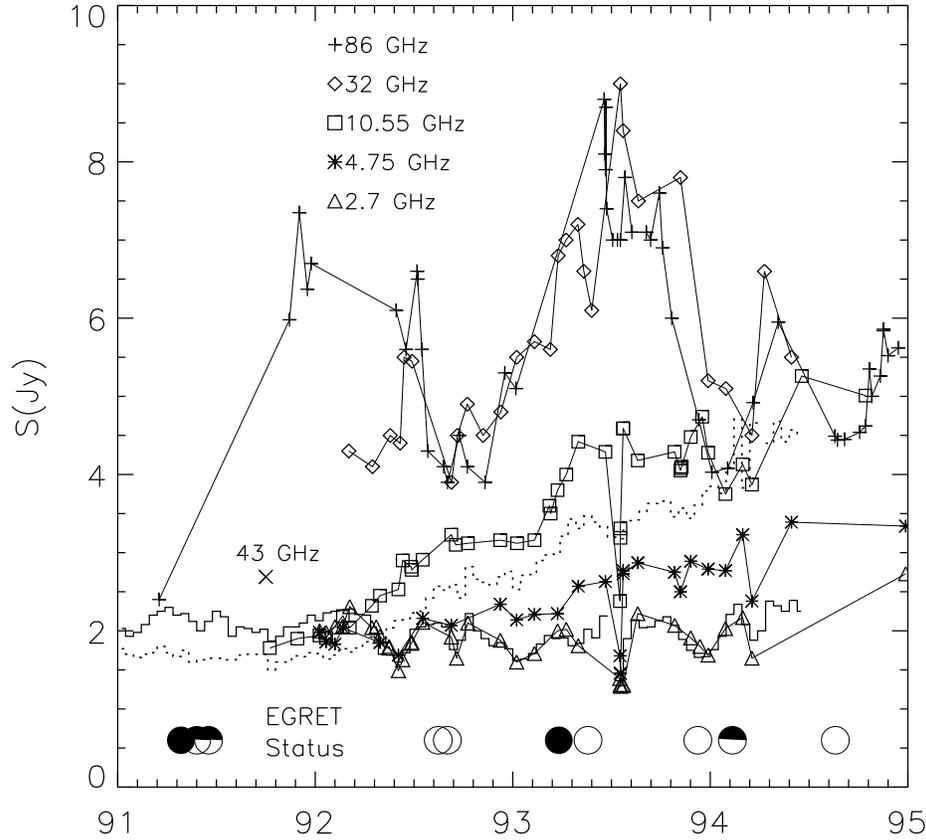

Figure 1. Radio lightcurves of 0528+134 from the Effelsberg 100-m, the IRAM 30-m and the NRL-GBI telescopes. The Green Bank data are given as averages over two weeks in histogram mode, where the full line denotes data at 2.25 GHz and the dotted lines give data at 8.3 GHz. The source activity in the EGRET range is indicated by open circles for low level, half-filled circles for medium activity, and filled circles for a high $\gamma$-ray flux.

## 2. VLBI observations of 0528+134

0528+134 was observed with a global VLBI array at 22 GHz in November 1992. Within the IRIS-S and EUROPE geodetic VLBI observing campaigns the source was also observed regularly at 8.4 GHz. The VLBI structure of 0528+134 at 22 GHz (epoch 1992.85) shows a one-sided core jet structure of ∼ 5 mas length, which is also typical for other epochs.

One of the components exhibits superluminal motion with an apparent velocity of $\beta_{app}$ = 4.4±1.7 (for $H_0$ = 100 km/s/Mpc, $q_0$=0.5). Superluminal motion in 0528+134 was expected since Doppler boosting is required to satisfy the compactness limit at MeV photon energies (McNaron-Brown *et al.*, 1995). Due to the strong reduction of the pair production cross section in the Klein-Nishina limit there is no similar requirement for γ-rays in the EGRET range (Pohl *et al.*, 1995).

Since only few EGRET sources have been observed with sufficient resolution in VLBI, the high percentage of superluminal sources in the sample of EGRET detected quasars yields strong evidence for the assertion that high energy γ-ray emission from AGN originates in relativistic jets and is strongly boosted around the direction of the jet.

A new VLBI component which is not present in data at 22 GHz from 1991 (Zhang *et al.*, 1994) also exhibits superluminal motion. Backextrapolating in time we find that the new component was expelled from the core in the first half of 1991, near the first γ-ray outburst in Mai 1991. Newest VLBI data indicate the appearance of another new component in 1994(Krichbaum *et al.*, 1995). Although this new component has yet been found only at two epochs, it appears to move superluminally and may have been expelled from the core between fall 1992 and summer 1993, i.e. near the time of the second gamma ray outburst and at the beginning of a strong radio outburst.

If the evidence for such a correlation can be solidified in future measurements, it would nicely complement to our view that activity in the central source of AGN results in the expulsion of plasma blobs (or shocks) at relativistic velocity, in which relativistic particles, most probably electrons, produce high energy gamma ray emission when the blob is still near to the core, and which after some cooling and expansion become visible at mm and cm wavelengths.